# Decoupling the Roles of Defects/Impurities and Wrinkles in Thermal Conductivity of Wafer-scale hBN Films

Kousik Bera[1], Dipankar Chugh[2], Aditya Bandopadhyay[3], Hark Hoe Tan[2,4], Anushree Roy[5], and Chennupati Jagadish[2,4]

[1]School of Nano Science and Technology, Indian Institute of Technology Kharagpur, Pin 721302. India.

[2]Department of Electronic Materials Engineering, Research School of Physics, The Australian National University, Canberra, ACT 2600, Australia.

[3]Department of Mechanical Engineering, Indian Institute of Technology Kharagpur, Pin 721302, India.

[4]Australian Research Council Centre of Excellence for Transformative Meta-Optical Systems, Research School of Physics, The Australian National University, Canberra, ACT 2600, Australia.

[5]Department of Physics, Indian Institute of Technology Kharagpur, Pin 721302, India.

## Abstract

We demonstrate a non-monotonic evolution of thermal conductivity of large-area hexagonal boron nitride films with thickness. Wrinkles and defects/impurities are present in these films. Raman spectroscopy, an optothermal non-contact technique, is employed to probe the temperature and laser power dependence property of the Raman active $E_{2g}^{high}$ phonon mode, which in turn is used to estimate the rise in the temperature of the films under different laser powers. As the conventional Fourier law of heat diffusion cannot be directly employed analytically to evaluate the thermal conductivity of these films with defects and wrinkles, finite element modeling is used instead. In the model, average heat resistance is used to incorporate an overall defect structure, and Voronoi cells with contact resistance at the cell boundaries are constructed to mimic the wrinkled domains. The effective thermal conductivity is estimated to be 87, 55, and 117 W/m.K for the 2, 10, and 30 nm-thick films, respectively. We also present a quantitative estimation of the thermal resistance by defects and wrinkles individually to the heat flow. Our study reveals that the defects/impurities render a much higher resistance to heat transfer in the films than wrinkles.

# 1. Introduction

With the increasing demand for closely packed miniaturized electronic devices, heat management for 2D materials is becoming critical in next-generation technology. The longevity and performance of devices, such as microchips, processors, and batteries, strongly depend on their operating temperature. Exceptionally high thermal conductivity, about 3080-5300 W/m.K, has been measured for graphene [1,2]. The value reduces to 850-3000 W/m.K [3–8] for the graphene layer grown by chemical vapor deposition (CVD). However, the high electrical conductivity of graphene imposes constraints on its use in thermal dissipation applications in devices [9]. Monolayer hexagonal boron nitride (hBN) films with relatively high thermal conductivity, 751 W/m.K [10], and large band gap, 5.6 eV [11] are potentially ideal as dielectric substrates [12–15]. Almost an order of magnitude enhanced carrier mobility is found for graphene devices on hBN than devices on $SiO_2$ [16]. The high thermal stability of the hBN films helps to protect air-sensitive metallic surfaces and 2D layered materials [17]. Due to the unique combination of high thermal conductivity, large band gap, and notable mechanical properties (which include high flexibility and stretchability) [18], thin hBN layers are extraordinary materials for a wide range of applications, from nano-electronics to the space industry [16,19,20].

A few studies have explored the thermal conductivity of hBN [10,21,22] using Raman spectroscopy. However, these studies are on small-sized exfoliated films [10,21], which find limitations in practical applications. It has been reported that by using metalorganic vapor phase epitaxy (MOVPE)/chemical vapor deposition (CVD), wafer-scale hBN films can be grown [23–26]. However, the unavoidable formation of defects, grains, grain boundaries, impurities, and wrinkles in these large-scale films [23–25,27,28] can affect their thermal conductivity. Moreover, a comparative study of the resistance caused by defects/impurities and wrinkles in the heat transfer in these wafer-scale hBN films is yet to be reported in the literature.

We employ Raman spectroscopy, as a non-contact opto-thermal technique, to measure the temperature in the film pumped by a laser with a Gaussian beam spot. The same experimental technique, along with the Fourier law of heat diffusion analytically, has been commonly used in the literature to study the thermal conductivity of single crystal/exfoliated 2D materials [29,30]. However, defects/impurities and wrinkles present in wafer-scale films constrained us to follow this approach. The objective of the present study is to find the effective thermal conductivity ($\kappa_{eff}$) of these large-scale hBN films by exploiting both experimentally measured temperatures of the films for known laser powers employing Raman spectroscopy and finite element modeling (FEM) for the heat transfer in the films. We further aim to decouple the relative thermal resistance rendered by defects/impurities and wrinkles individually to the heat flow in these films. One of the main advantages of the FEM is that it studies the system at a macroscopic scale which is more suitable than molecular dynamics simulation (MD) or density functional theory (DFT) to investigate inhomogeneous 2D layers. Though one obtains the material's response at an atomic scale by MD/DFT approach, these methodologies demand a high computational cost to model a macroscopic inhomogeneous material, such as 2D films with defects, impurities, and wrinkles. In the present study, FEM considers the distribution of these structural parameters over the entire area of the film to explore the effective thermal properties of the 2D layers.

## 2. Experimental Details

The hBN films were grown on commercially available 2″ diameter sapphire substrates by a closed couple shower head type MOVPE reactor (AIXTRON). Growth was carried out at a temperature of 1350 $^{o}$C using triethylboron and ammonia as the precursors. The details of the growth process are available in Ref. [24]. The thickness of the films were determined using depth-profiling atomic force microscopy [24]. In this work, we used hBN films of thicknesses 2, 10, and 30 nm. The 30 nm films could be delaminated from the sapphire substrate spontaneously by bringing it in contact with de-ionized water. For the 2 and 10 nm-thick hBN films, spin-coating with PMMA followed by immersion in a 2% hydrofluoric acid bath was used for film transfer. Acetone was used to remove PMMA residue. The delaminated films were then rinsed and transferred onto (i) SiO$_2$ (2 µm) coated and (ii) gold (200 nm)/chromium (10 nm) coated prefabricated (using photo- and electron beam lithography) circular microwells of 3 µm diameter with a well-defined boundary. Here we would like to mention that the films were also suspended on microwells of larger diameters (8, 15, and 20 µm). However, the sagging of the films over microwells of these larger diameters constrained us to use only the wells of smaller diameters in further measurements (Fig. S1). Reports in the literature suggest that it is non-trivial to remove PMMA completely from films [31]. In view of this, we recorded Raman spectra of the suspended films over a wide spectral range and at many points on the suspended part of the film surfaces to trace the presence of residual PMMA. Fig. S2 plots the characteristic Raman spectra on 2 nm and 10-thick films at five different points. For further study, only the suspended films without the characteristic spectral signatures of PMMA were chosen. Earlier, it was shown [24] that carbon is the major impurity, which is incorporated in these films during the MOVPE growth process. Residual PMMA, if any, in 2 nm and 10 nm films is the additional source of impurity, which is absent in the 30 nm-thick samples. As the thermal conductivity of PMMA is low (0.167-0.25 W/m.K [32]), even if present, it may not affect the measured thermal conductivities of the 2 and 10 nm films appreciably.

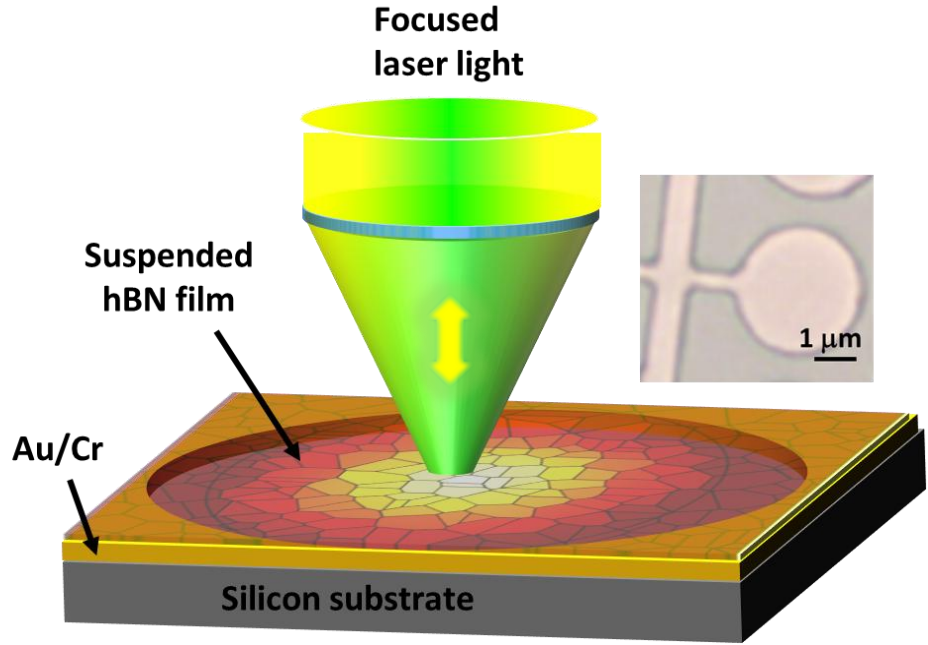

**Figure 1.** Schematic diagram of the experimental arrangement to measure the thermal conductivity of the films (not to the scale). Inset: Optical image of the fabricated microwells with trenches.

A schematic diagram of the experimental arrangement is shown in Figure 1. The optical image of one of the circular microwells with trenches, used in the experiment, is shown in the inset of Figure 1. During heating, the trenches allowed the trapped air in the hBN-covered microwell to escape rather than resulting in volume expansion. While the films suspended on $SiO_2$/Si microwells were used to study the temperature coefficient of the Raman shift of the $E_{2g}^{high}$ mode in these films, the ones suspended on Au-Cr/Si microwells were used to estimate the Raman shift of the same mode by laser radiation of known power. Atomic force microscopy (AFM) (Multimode 8, Bruker) under tapping mode operation was used for studying the surface topography of the films.

Raman measurements were performed with LabRAM HR Evolution (Horiba, France) system, which included a confocal microscope with a 100× objective lens. The microscope objective served for both focusing the incident laser light and collecting the back-scattered signal. 532 nm emission line of a diode laser was used as the excitation wavelength. A charge-coupled device (Syncerity, Horiba, France) was used to record the spectrum. The laser power on the sample was varied using neutral density filters in the path of the incident laser beam. Temperature-dependent Raman spectra of hBN were measured by placing the sample inside a cooling-heating microscope sample stage (THMS600, Linkam Scientific Inc. UK) and using a 50×L objective lens.

## 3. Results

### 3.1. Surface topography of the hBN films

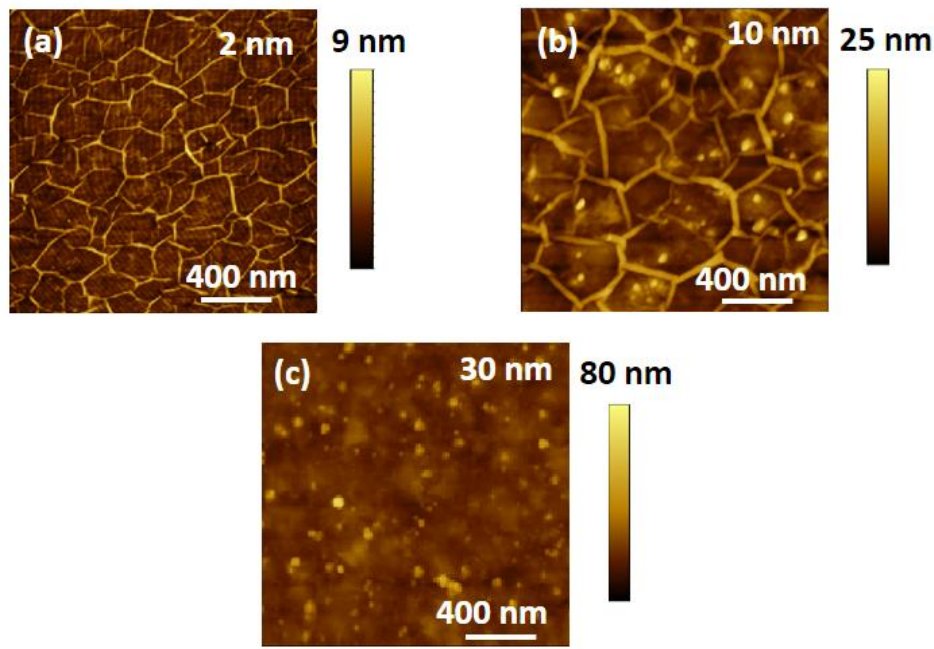

**Figure 2. AFM images showing the surface topography of the delaminated (a) 2 nm, (b) 10 nm, and (c) 30 nm-thick hBN films, for which the thermal conductivity has been measured.**

Defects and wrinkles are present in wafer-scale MOVPE-grown films. Detailed structural characteristics of as-grown films, obtained from Raman mapping, PL, cross-sectional transmission electron microscopy, x-ray photoelectron spectroscopy, and atomic force microscopy are available in Ref. [24, 33]. During cooling after the growth in the reactor chamber, the film experienced a compressive strain which led to wrinkling to release the energy due to the difference in lattice thermal expansion coefficient between the film and the sapphire substrate on which it is grown [24, 33].

Figure 2 presents the surface topography of the delaminated 2, 10, and 30 nm-thick hBN films, on which the thermal conductivity measurements are carried out. It is to be noted that the wrinkles remained in the 2 nm, and 10 nm-thick delaminated films. The delaminated 30 nm-thick film is free from wrinkles. Photoluminescence (PL) is often used to characterize defects/impurities in these hBN films [24]. Because of structural inhomogeneity, the PL spectra were recorded at many points on a film surface. Characteristic PL spectral profiles in Fig. S3 of all three films are not significantly different, considering the extent of defects/impurities presents in them. It is to be mentioned here that the spectra recorded at all points for a particular film thickness were not identical. However, the overall spectral profile remained the same.

**3.2 Estimation of the temperature coefficient of Raman shift**

Over the last couple of decades, Raman spectroscopy has emerged as a precise contactless technique to measure the thermal conductivity of 2D materials [34-37]. The focussed laser beam, which is used as the excitation source in Raman measurements, locally generates heat. By quantifying the intensity ratio of the Stokes ($I_S$) and anti-Stokes ($I_{AS}$) modes from Raman measurements, one can deduce the temperature rise in the sample due to absorbed laser power [35]. From this measured parameter, one can estimate the thermal conductivity of the sample under investigation. Unfortunately, the intensity of the in-plane phonon mode of hBN at ~1367 cm$^{-1}$ is very weak in non-resonant Raman spectroscopy, and therefore, it is non-trivial to record anti-Stokes Raman signals of hBN. Thus, the ratio $I_S/I_{AS}$ could not be used to estimate the temperature of the thin hBN films for the available laser powers.

We followed an indirect approach to estimate the temperature rise in the films under laser radiation. First, we calibrated the Raman shift of the given hBN films for various applied temperatures keeping laser power fixed at a low value of 0.20 mW, so that the additional heating of the samples due to the laser radiation is negligible. To find the temperature coefficient of Raman shift, temperature-dependent Raman spectra of three wafer samples suspended over SiO$_2$/Si microwells were recorded. Figure 3(a)–(c) display the temperature dependent Raman spectra of 2, 10, and 30 nm hBN films, respectively, suspended over SiO$_2$/Si microwells over the spectral range between 1315 and 1425 cm$^{-1}$ (a magnified view of the plots showing the shift with temperature is shown in Fig. S4). Each spectrum was fitted with a Lorentzian function, with the intensity, width, and peak position as the free fitting parameters. The relative Raman shifts ($\Delta\omega=\omega(T)-\omega(297K)$, where $\omega(T)$ and $\omega$ (297K) are the measured Raman shift at temperature $T$ and at 297K, respectively) with temperature for the three different samples, summarized in Figure 3(d), show a downward shift with increasing temperature for all samples.

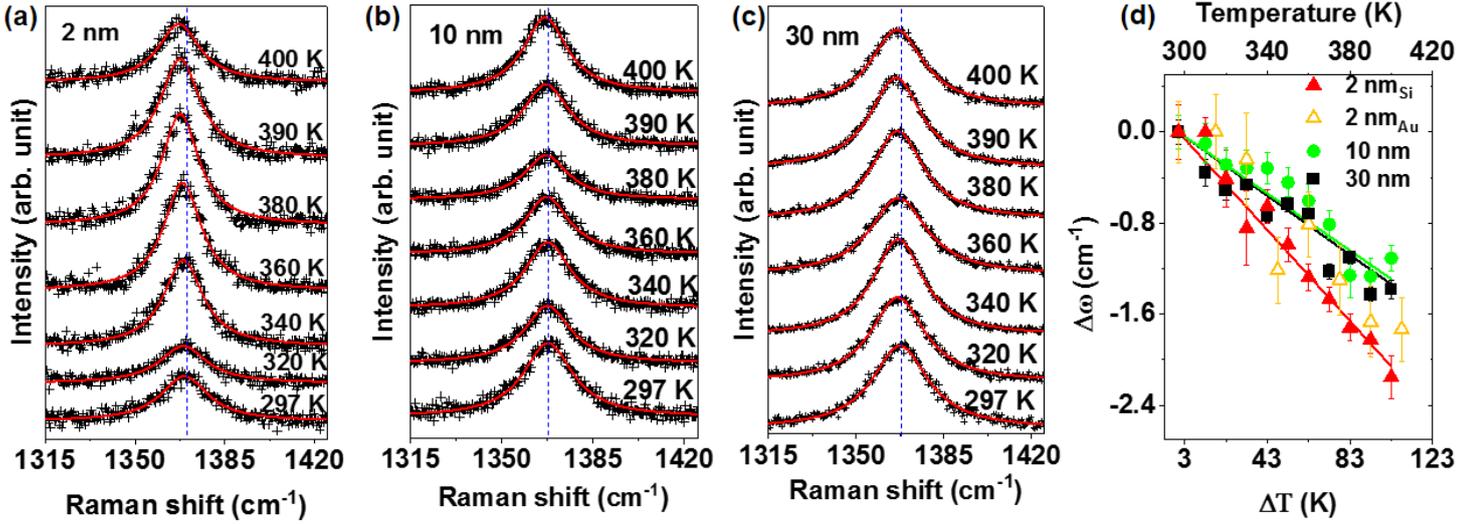

**Figure 3.** (a)–(c) Characteristic Raman spectra of $E_{2g}^{high}$ mode of 2, 10, and 30 nm-thick films suspended over SiO$_2$/Si microwells at different heating temperatures. The blue dashed lines mark the position of the $E_{2g}^{high}$ mode peak at 297K with temperature. The zoomed view is available as Fig. S4. All spectra are not recorded with the same integration time. (d) Relative change in Raman shift ($\Delta\omega$) of the $E_{2g}^{high}$ mode (from its room temperature value) as a function of change in temperature ($\Delta T$) and the corresponding linear fit to the data points. Red filled triangles, green circles, and black squares correspond to the data points for the 2, 10, and 30 nm-thick films, respectively. Gold open triangles are for the 2 nm film on Au-Cr/Si microwells. The top axis labels the set values of the temperature.

Various factors are responsible for the evolution of the Raman shift of the suspended 2D films with temperature. Details of the thermal effect on the Raman shift of suspended hBN films have already been discussed by us in Ref. [38]. For large-scale hBN films, the thermal contraction of the lattice and the increase in anharmonicity with the rise in temperature can play a dominant role. Additionally, the contact regions of films with the side wall of the microwells could give rise to strain due to thermal expansion coefficient (TEC) mismatch. Figure 3(d) also includes the variation of $\Delta\omega$ with temperature for the 2 nm-thick film suspended on a Au-Cr/Si microwell. As the TEC mismatch between hBN and the substrate is more for Au-Cr/Si than for $SiO_2$/Si [39], it is expected that the effect of differential TEC would have been more prominent in the former than for films on $SiO_2$/Si-coated microwells. However, the data points for both sets of samples (open and closed triangles) nearly overlap, as shown in Figure 3(d). It suggests that the contact regions of the film with the substrate play an insignificant role on $\Delta\omega$. The first order temperature coefficient of Raman shift ($\chi$) could be estimated by assuming a linear relation [10], i.e., $\Delta\omega=\chi\Delta T+c$, where $c$ is the $y$-intercept. From the linear fit to the data points, the values of $\chi$ ($=d(\Delta\omega)/d(\Delta T)$) are estimated to be $-0.02\pm0.001$, $-0.010\pm0.002$, and $-0.010\pm0.002$ cm$^{-1}$/K for the 2, 10, and 30 nm-thick suspended films, respectively.

### 3.3 Evolution of Raman shift with temperature rise due to applied laser power

Next, the incident laser beam was used as the heat source on the suspended films. The beam was focused at the center of the film on the microwell. The rise in temperature due to the laser power was investigated by recording Raman spectra of the films suspended over Au-Cr/Si microwells. The excitation power-dependent Raman spectra of the films over the spectral range between 1345 and 1400 cm$^{-1}$ are shown in Figure 4 (a)-(c). The power level indicated next to each spectrum is the measured (applied) power on the sample. The change in Raman shift due to applied laser powers is shown in Figure 4(d). The Raman mode downshifts for all the cases, which indicates an increased local temperature with increasing laser power. Here the change in Raman shift due to laser power relates to the thermal conduction of hBN films towards the Au-Cr/Si heat sink. By using the temperature coefficient of Raman shift ($\chi$), the temperature ($T_m = d(\Delta T)+297$) of the film as a function of applied laser power is estimated by following the relation

$$d(\Delta T) = \frac{d(\Delta \omega)}{dP} \cdot \frac{d(\Delta T)}{d(\Delta \omega)} dP \qquad (1)$$

and will be discussed later.

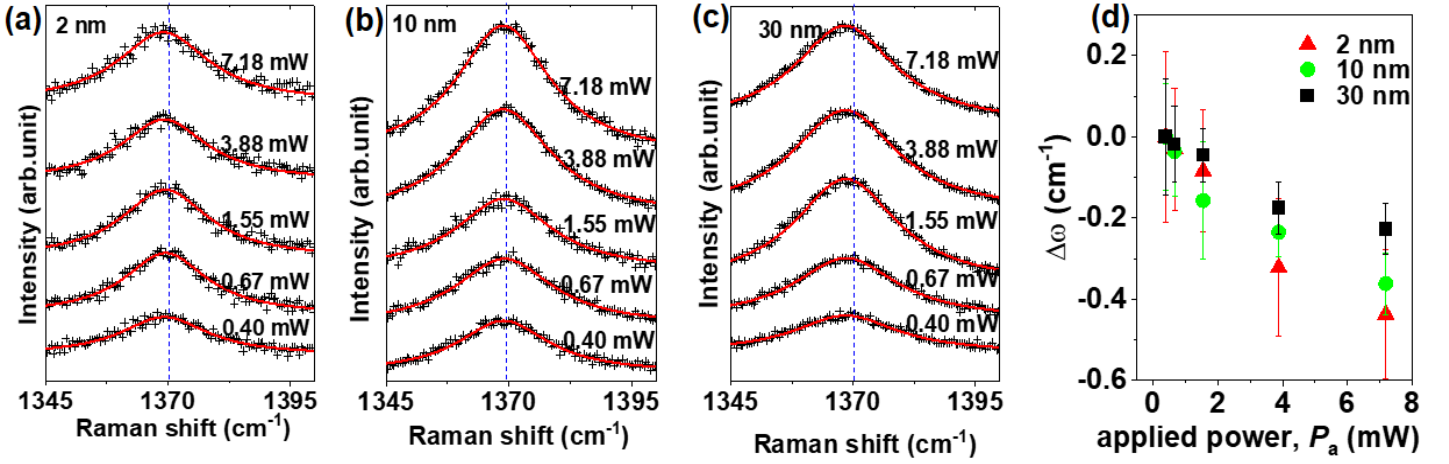

**Figure 4.** (a)-(c) Raman $E_{2g}^{high}$ mode of the 2, 10, and 30 nm-thick hBN films suspended over Au-Cr/Si microwells at different laser powers. + symbols are measured data points. Each spectrum is fitted with a single Lorentz function (solid lines). Blue dashed lines in each panel marks the position of the $E_{2g}^{high}$ mode peak at 297K. (d) Change in $E_{2g}^{high}$ mode frequency as a function of laser power.

## 4. Discussion

### 4.1 Finite element modeling for the heat transfer

For a homogenous film, the temperature distribution $T(\vec{r})$ across the film due to a heat source $Q(\vec{r})$ is governed by the steady state heat diffusion equation

$$\kappa \cdot \nabla^2 T(\vec{r}) + Q(\vec{r}) = 0, \tag{2}$$

where $\kappa$ is the thermal conductivity of the material. For a Gaussian laser beam as the isotropic heat source and its incidence at the centre $|\vec{r}_0|$ of the suspended part of the film,

$$Q(\vec{r}) = \frac{P}{\pi d a^2} \exp\left(-\frac{|\vec{r} - \vec{r}_0|^2}{a^2}\right). \tag{3}$$

Here, $a$ is the radius of the laser spot, and $P$ is the absorbed laser power. $d$ is the thickness of the film. Introducing a dimensionless variable $\vec{\rho} = \vec{r}/a$, the solution of Eqn. 1 for a homogeneous film and the isotropic heat source is

$$T(0) - T(\rho) \approx \frac{\alpha}{2}\left(\ln \rho + \frac{\gamma}{2}\right), \tag{4}$$

with $\alpha = P/\pi \kappa d$ and $\gamma$ is the Euler's constant. $T(0)$ is the temperature of the film at the center of the beam. In our experiments (for example, as shown in Figure 1), the edge of the suspended film can be considered to be at ambient temperature, i.e., $T(R/a)=297$ K, where $R$ is the radius of the microwell. Note that the parameter $P$ in Eqn. 2 is not the same as the applied laser power $P_a$ (in Figure 4), as only part of the incident radiation is absorbed by the film. Furthermore, some of the absorbed power may be lost by air convection and radiation. For a homogeneous film, one can estimate the loss due to these factors analytically [5,10]. By measuring $T(0)$ with respect to the ambient temperature at the edge, one can estimate $\kappa$ [35]. However, for a film with grain boundaries, defects, impurities, or wrinkles, an additional damping term must be included in the LHS of Eqn. 1 [36]. Nonetheless, in this way, the role of defects/impurities and wrinkles cannot be decoupled. Furthermore, it is non-trivial to estimate the heat loss by the radiation for these inhomogeneous films.

Thus, instead of estimating the thermal conductivity of the film using the above analytical formalism, we determine it numerically using finite element approach. This method describes a system at a macroscopic scale and finds approximate solutions of partial differential equations numerically. The solutions are based on subdividing a large system into smaller parts and linearizing the equations within each of the small parts. The heat propagation in hBN films is simulated using the heat transfer module in COMSOL Multiphysics (version 4.4) software package. Using this module, temperature across the film can be mapped by considering the characteristics of the films and the heat-generating source.

To mimic the experimental geometry of the suspended film in the simulation, we define a circular boundary of radius $R$ for the film. The modeled hBN films of known thicknesses (2 nm, 10 nm, or 30 nm) and specific heat are subjected to a Gaussian laser beam of radius $a$ at the center (as in Eqn. 3). The heat distribution due to the laser power takes place within the circular boundary, beyond which room temperature is maintained. The model assumes an ideal thermal contact between the film and the region beyond the boundary (since we used Au-Cr/Si as the heat sink in the experiment). In the simulation, we use the values of $R$ and $a$ as 1.5 μm and 0.5 μm, respectively, corresponding to the values determined experimentally. UV-visible spectroscopy was carried out to find the absorbance at 532 nm by the films. The absorbance values for the 2, 10, and 30 nm films are 0.23%, 1.15%, and 3.4%, respectively. In the simulation, the applied laser power $(P_a)$ is scaled by the above absorbance values for the films to obtain $P$. Heat loss to the air, which is the sum of radiative and convective components, is also included as

$$Q_{air} = h(T - T_{amb}) + \varepsilon\sigma(T^4 - T_{amb}^4),$$

where, $h$, and $\varepsilon$ are heat transfer coefficient (3475 W/m$^2$.K) and surface emissivity (0.8) for hBN, respectively, $\sigma$ is the Stefan-Boltzmann constant, and $T_{amb}$ is the ambient temperature (297 K) [22]. Since Raman measurements are carried out in back-scattering geometry, we obtain an average value of the temperature over the area of the laser spot size, see Fig. 1. Thus, in the simulation, we also use the mean temperature ($T_m$) within the spot size of the laser beam of radius $a$. In order to minimize the error in computation, the 'extremely fine' meshing in COMSOL is used for FEM.

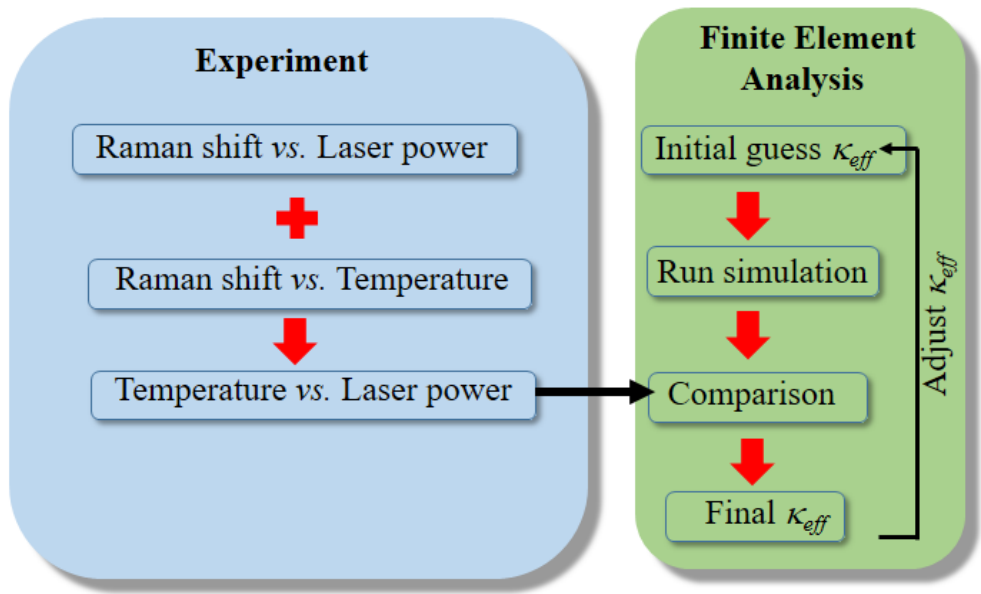

**Figure 5. Combined experimental and simulation workflow to obtain effective thermal conductivity of the films.**

For suspended films in the presence or absence of wrinkles, defects, and impurities, their effective thermal conductivity ($\kappa_{eff}$) determines the rise in temperature due to heat flux under different laser powers. In the simulation, for a given modeled film, the value of $\kappa_{eff}$ was adjusted to obtain the measured values of $T_m$ at different laser powers. It is to be noted that a homogenous film with a higher value of $T_m$ effectively models the film in which heat flow finds higher resistance. The workflow to obtain $\kappa_{eff}$ of a film, using the experimental results obtained from Raman measurements and FEM of the heat flow is summarized in Figure 5.

## 4.2 Estimation of the effective thermal conductivity ($\kappa_{eff}$) of the films

To estimate $\kappa_{eff}$, Figure 6 plots the measured temperature $T_m$ of the 2, 10, and 30 nm-thick films at different applied laser powers (as red triangles, green circles, and black squares, respectively) using the results obtained in Fig. 4(d) and the value of $\chi$ for each film thickness in Eqn. 1.

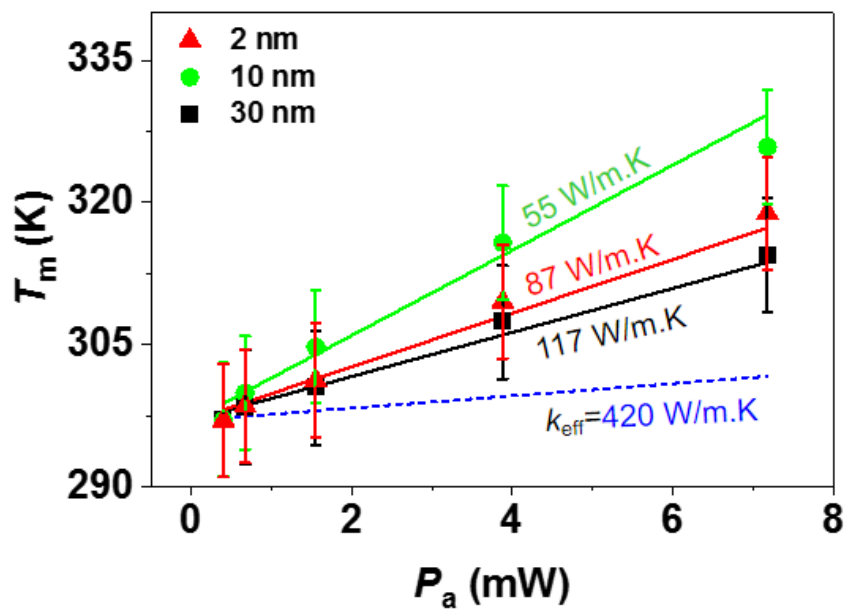

**Figure 6. Measured temperatures for different applied laser powers for films of different thicknesses (data points). The solid lines are simulated for $T_m$ with the appropriate values of effective thermal conductivity ($\kappa_{eff}$) that fit the data points for each film thickness. The dashed line shows the expected change of $T_m$ with applied laser power for bulk hBN.**

We simulate the temperature ($T_m$) of the films of different thicknesses for different values of $\kappa_{eff}$. The experimental data points could be best matched (solid lines in Figure 6) for the value of $\kappa_{eff}$ as 87, 55, 117 W/m.K for the 2, 10, and 30 nm films, respectively. The dashed line in Figure 6 plots the variation of $T_m$ with applied laser power for a $\kappa_{eff}$ value of 420 W/m.K of bulk hBN [40]. For a homogeneous hBN film, one expects a systematic reduction in the value of the thermal conductivity with an increasing film thickness upto a few layers [10] and reaches the same value as that of bulk hBN [41]. The above estimated values of $\kappa_{eff}$ for the 2, 10, and 30 nm films with the approximate number of layers of 6, 30, and 90, respectively (with a thickness of each layer as 0.33 nm [24]), are significantly less compared to that of bulk hBN due to the presence of defects/impurities and wrinkles. Moreover, we find that the value of $\kappa_{eff}$ does not change monotonically with the thickness of these films. As shown in Figure 2, unlike the 2 and 10 nm films, the 30 nm film is free of wrinkles. Defects/impurities are, however, present in all samples. Furthermore, the density of wrinkled domains is more for the 2 nm film than the 10 nm film (See Figure 2). The non-monotonic trend in $\kappa_{eff}$ with thickness indicates a complex role of wrinkles and defects/impurities in the films, which we discuss below.

**4.3 Thermal resistance by defects and impurities in hBN films**

In Figure 6, it is interesting to note that the presence of defects/impurities reduces the value of $\kappa_{eff}$ of the 30 nm film by nearly 70% from its value (420 W/m.K [39]) in bulk hBN. To estimate the thermal resistance caused by defects/impurities, the model 30 nm-thick film (free from wrinkles) was exposed to a Gaussian laser beam of different laser powers, as above. A mean thermal resistance ($R_D$) over the suspended area of the model film is introduced in the simulation to model the resistance to the heat flow caused by defects in the film. The value of $R_D$ was varied to obtain the contour plot of the simulated temperature $T_m$ for different laser powers (refer to Figure 7(a)). The dashed lines are isothermal lines for the experimentally measured values of $T_m$ for the 30 nm-thick film under different laser powers (from Figure 6). For a given laser power (on the y-axis), a point on an isothermal line yields the temperature of the film for a particular value of $R_D$ (on the x-axis of the plot). For $R_D = 14$ nK.m$^2$/W, the simulated variation of $T_m$ with applied power matches the experimentally obtained results in Figure 6 (see diamond symbols). It is to be noted that the estimated value of $R_D$ is the heat resistance in the film only due to disorders/impurities.

### 3.5 Thermal resistance due to wrinkles in hBN films

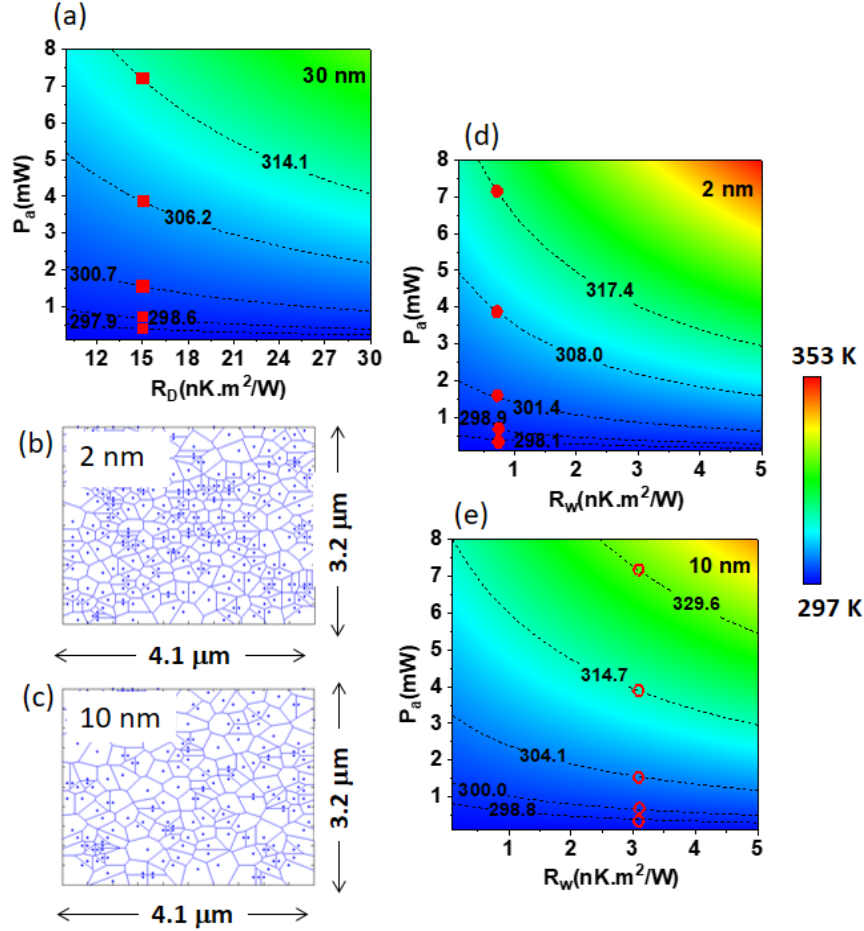

**Figure 7.** (a) Contour plots of $T_m$ for different laser powers and $R_D$ for the 30 nm-thick film. The dashed lines are isothermal plots for the experimental values of $T_m$. The square symbols indicate the values of $R_D$ on the isothermal lines for the applied laser powers used in the experiments. Models of wrinkled hBN films for FEM simulation: (b) average domain size of 150 nm, and (c) average domain size of 260 nm. (d) and (e) Contour plots of $T_m$ for different laser powers and $R_W$ for the 2 and 10 nm-thick films, respectively. The dashed lines are isothermal plots for the experimental values of $T_m$. The filled and open symbols indicate the values of $R_W$ on the isothermal lines for the applied laser powers used in the experiment for 2 nm and 10 nm-thick films, respectively.

As mentioned above, in view of the fact that the PL spectra of all three films are also very similar [see Fig. S3], we assume that defects/disorders in these films are not significantly different. Nonetheless, in addition to defects/impurities, wrinkles are also present in the 2 and 10 nm films. However, the density of wrinkles is significantly different in these two films. To model the 2 and 10 nm films, a discrete set of points (seeds) are randomly distributed across a plane with a predefined dimension (4.1 μm × 3.2 μm). Based on the distributed points, Voronoi cells are constructed (refer to Figure 7(b) and (c)). The Voronoi cell has the property that any point X in the plane within the cell R(Y) is closer to seed point Y than any other seed point. The average Voronoi cell size is kept close to the measured wrinkle wavelengths of 150 and 260 nm to represent 2 nm and 10 nm films, respectively. There are typically 160 and 100 domains in the 2 nm and 10 nm modeled films, respectively.

Quite a few articles in the literature have discussed the role of wrinkles in determining the thermal conductivity of graphene [36,42]. Non-equilibrium molecular dynamics simulation [43] suggests that the strong localization of phonon across the wrinkled domain wall reduces the thermal conductivity of the film. In view of this, in addition to the 14 nK.m$^2$/W value used for $R_D$ to represent the effect of defects/impurities, resistance ($R_w$) as contact elements of domain boundaries is also introduced. In our model, we assume that defects/impurities and wrinkles render independent resistive processes [44]. As before, a circular boundary of radius $R$ of the modeled film is subjected to the radiation of a Gaussian laser beam of radius $a = 0.5$ μm at the center of the film. Keeping other constraints the same as described above, Figure 7(d)-(e) present the simulated contour plots of $T_m$ for different applied laser powers and $R_w$ using a false color scale for the 2 and 10 nm films. The isothermal lines in Figure 7(d) and (e) are shown by dashed curves for the experimentally measured temperatures $T_m$ of the films obtained in Figure 6. The filled and open symbols in panels (d) and (e) on the isothermal lines mark the value of $T_m$ at values of $P_a$ used in the experiments for 2 nm and 10 nm-thick films, respectively. The corresponding values of $R_w$ are 0.7 and 3 nK.m$^2$/W, respectively, for the 2 and 10 nm films, respectively. The resistance per unit thickness due to the wrinkled domain boundaries is ~0.35 (0.3) nK. m/W for the 10 nm (2 nm)-thick film. In the above analysis, we find that the resistance to the heat flow caused by defects/impurities is much higher than that caused by wrinkles.

**3.6 Surface temperature distribution**

Furthermore, we use FEM to simulate the surface temperature profile over the suspended area for the given experimental geometry and laser induced heating at the center of the films. Fig.

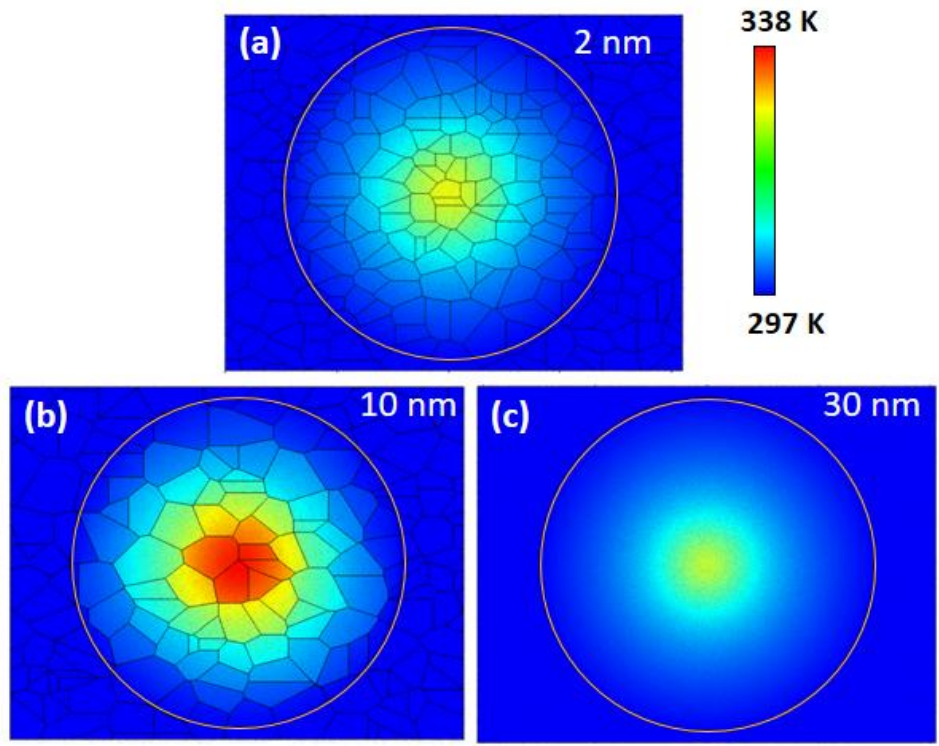

**Figure 8 (a)-(c) Simulated surface temperature distribution of the 2, 10, and 30 nm suspended films, respectively, under laser heating. The suspended part in each panel is within the golden circular boundary.**

8(a)-(c) plot the profiles for the 2 nm, 10 nm, and 30 nm-thick films, under study. The least value of $\kappa_{eff}$ in the 10 nm film compared to that of the 2 nm and 30 nm films (as estimated earlier), results in a maximum rise in temperature for the 10 nm-thick film. It would be interesting to experimentally verify the above non-monotonic trend in temperature distribution in the large-scale films, which could not be carried out in the present study due to the sagging of the films mounted over larger diameter microwells. Such information would be important for applications of these films in the thermal management of 2D material based devices.

## 5. Summary

We carry out a combined experimental and modeling approach to study the effective thermal conductivity of wafer-scale hBN films of different thicknesses. Defects/impurities and wrinkles in these large area films reduce the thermal conductivity from their values expected in homogeneous counterparts drastically. The present study aims to decouple the contribution of defects/impurities and wrinkles in determining the effective thermal conductivity of the films. Raman spectroscopy, a non-contact technique, is used for measuring the temperature of the film for different laser powers. The measurements are performed on 2, 10, and 30 nm-thick suspended hBN films. Due to the inhomogeneity of the films, instead of solving the Fourier diffusion law analytically, we apply FEM to estimate the effective thermal conductivity of the films by including wrinkles and defects/impurities in the models. To model the wrinkles, Voronoi cells are constructed following the characteristic morphology of the measured wrinkled domains in the 2 and 10 nm films. The combined experimental and simulated observations suggest that (i) non-monotonic change in the thermal conductivity of the films with thickness, and (ii) the resistance to heat flow caused by defects/impurities is much higher than that caused by wrinkles. Though the present work is on MOVPE-grown hBN films, the same methodology described here can be adopted to find $\kappa_{eff}$, and the relative role of defects/impurities and wrinkles in any other 2D system fabricated by other techniques, like MBE and CVD. Further, the study shows that defects/impurities, wrinkles, and thickness in hBN films can be exploited to modulate the phonon density of states and hence, to use the system efficiently for thermal management and thermoelectric devices.

**Acknowledgment:**

We acknowledge funding from the Australian Research Council. ANFF ACT Node is acknowledged for providing access to the facilities used to grow the hBN samples.